\documentclass[a4paper, oneside, twocolumn, notitlepage, 10pt]{extarticle_ecoc}
\usepackage{ecoc}
\usepackage{amsmath}
\usepackage{tikz}
\usepackage{pgfplots}
\usepackage{mathtools}

\pgfplotsset{compat=1.18}
\usepgfplotslibrary{colormaps}
\usepgfplotslibrary{colorbrewer}
\usepgfplotslibrary{groupplots}
\usepgfplotslibrary{external} 
\usetikzlibrary{arrows.meta}
\usetikzlibrary{arrows}
\usetikzlibrary{fpu}
\usetikzlibrary{shapes.geometric}
\usetikzlibrary{external}
\usepackage{xpatch}

\makeatletter
\patchcmd\blx@bblinput{\blx@blxinit}
                      {\blx@blxinit
                      }{}{\fail}
\makeatother

\definecolor{grad-1}{HTML}{C7B3CC}
\definecolor{grad-2}{HTML}{A7ABC7}
\definecolor{grad-3}{HTML}{87A3C2}
\definecolor{grad-4}{HTML}{669ABC}
\definecolor{grad-5}{HTML}{4692B7}
\definecolor{grad-6}{HTML}{268AB2}

\begin{document}
\selectlanguage{english}    

\title{Ultrawideband optical fibre throughput in the presence of total optical power constraints from C to OESCLU spectral bands}

\author{
    Mindaugas Jarmolovi\v{c}ius\textsuperscript{(1)},~Henrique Buglia\textsuperscript{(1)},~Eric Sillekens\textsuperscript{(1)},~Polina Bayvel\textsuperscript{(1)},~Robert I. Killey\textsuperscript{(1)}}

\maketitle                  

\begin{strip}
    \begin{author_descr}

        \textsuperscript{(1)} Optical Networks Group, UCL (University College London), London, UK,
        \textcolor{blue}{\uline{zceemja@ucl.ac.uk}}
    \end{author_descr}
\end{strip}

\renewcommand\footnotemark{}
\renewcommand\footnoterule{}

\begin{strip}
    \begin{ecoc_abstract}
Using a recently developed fast integral ultrawideband Gaussian noise model, we quantify the achievable throughput under total optical power constraints for systems ranging from C-band to fully populated OESCLU bands using optimum launch powers, showing conditions when expanding bandwidth provides no additional throughput.
        ~\textcopyright2024 The Author(s)
    \end{ecoc_abstract}
\end{strip}

\section{Introduction}

Future demands for data traffic can be met through the exploitation of a greater fraction of the usable bandwidth within optical fibres, defined by the low-attenuation window of silica, across the O- to U-band (1260~nm to 1675~nm), as shown in Fig.~\ref{fig:fibre}. Increasing the transmission bandwidth to ultrawideband (UWB) is an attractive and cost-effective approach to meet the ever-growing traffic demand through the reuse of existing deployed fibres ~\cite{van_den_hout_ultra-wideband_2024}.

Currently, deployed systems mostly operate over narrow bandwidths using the C or C+L bands. However, the use of other available bands has been assessed in experimental field trials, which have achieved milestones in data throughput. Landmark UWB transmission exeriments have already been demonstrated to cover O- to U-bands \cite{soma_25-thz_2023,puttnam_301_2023}, reaching rates as high as 379~Tbps after 50~km~\cite{OFC24_378}. With recent advances in system design \cite{mikhailov_12551355_2024,donodin_pump_2023} and transmission~\cite{elson_continuous_2024,hazarika_multi-band_2024,donodin_multi-band_2024} with bismuth-doped fibre amplifiers (BDFA), we expect this achievable data rate to increase further.

However, network operators often have very conservative constraints on optical powers over installed networks in consideration of possible equipment and power rating limits in optical switches, couplers, and poor fibre splices; as well as eye safety regulations, as total optical power can exceed that of Class 4 lasers. At high optical power, fibres are also at risk of failure when under excessive bending \cite{glaesemann_analysis_2004}. 

In this work, for the first time, we analyse the maximum achievable throughput using the optimum launch power profile of short- and long-haul optical systems ranging from C-band only (4.38~THz) to O- to U- bands (58.95~THz) under the conditions of unconstrained and constrained total launch power. We quantify the impact of total power constraints on the overall net throughput achievable with launch power optimisation using a new integral GN model.  Although the inter-channel stimulated Raman scattering Gaussian Noise (ISRS GN) model is a commonly used tool for the rapid estimation of fibre channel nonlinear interference (NLI) noise ~\cite{isrsgnmodel} and has been used to estimate system performance using the E- to U-bands~\cite{poggiolini_closed_2022} together with the O-band~\cite{uwb_lidia,zefreh_accurate_2021}, the first demonstration of its accuracy in integral form for the fully loaded O-band and all the other bands, including ISRS, has recently been reported in~\cite{jarmolovicius_optimising_2024}, allowing for the first time estimation of the full O- to U-bands transmission performance. 

\section{Model and system setup}

\begin{figure}[t!]
    \vspace{-2em}
    \centering
    \clearpage{}\includegraphics{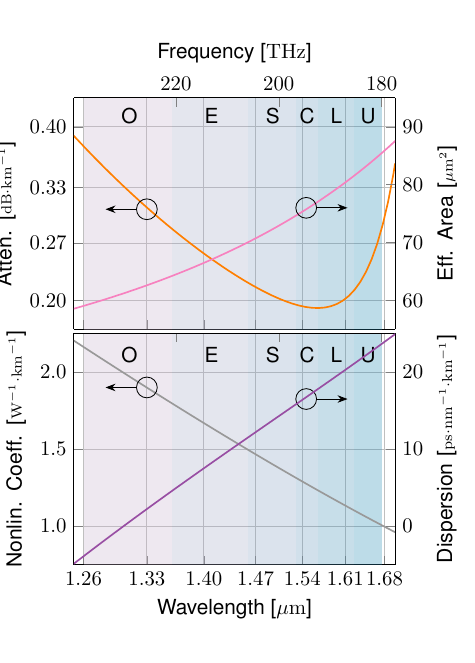}
\clearpage{}
    \vspace{-4em}
    \caption{Fibre attenuation, effective core area, nonlinear coefficient, and dispersion.}
    \label{fig:fibre}
\end{figure}

Taking into account the next-generation transceiver technology, we simulate a transmission system where each channel is modulated at $f_s=148~\text{GBaud}$ symbol rate with 150~GHz channel spacing. We use $L=80$~km span length standard single mode fibre taking into account wavelength-dependent attenuation, nonlinear coefficient, effective area and dispersion as shown in Fig.~\ref{fig:fibre}. Single- and 6-span transmission with Gaussian constellations and ideal lumped amplification are considered in the modelling. We consider several benchmarking scenarios of 15~dBm, 20~dBm and 25~dBm and an unconstrained upper limit of total optical launch power. Moreover, simulations are carried out assuming both ideal noise-free transceivers, and noisy transceivers with back-to-back SNR = 20~dB.
We assume a power leveller in the transmitter and after each in-line lumped amplifier, such that we can achieve an arbitrary spectral launch power profile after each span. We use realistic values for amplifier performance, the constant per band noise figure (NF)~\cite{poggiolini_closed_2022} as shown in Tab.~\ref{tab:params}.

We employ the optimised integral ISRS GN model~\cite{jarmolovicius_optimising_2024}, to accurately estimate nonlinear interference (NLI) noise in O- to U-band systems. 
To ensure high accuracy, we use $N_R=150$ and $\bar{N}_M=1.4$ as model parameter values.

\begin{table}[t!]
    \centering
    \caption{Amplifier noise figure (in dB), number of channels and number of segments per band used in the model.} \label{tab:params}
    \begin{tabular}{|c|c|c|c|c|c|c|}
        \hline Band & \cellcolor{grad-1!50}O & \cellcolor{grad-2!50}E & \cellcolor{grad-3!50}S & \cellcolor{grad-4!50}C & \cellcolor{grad-5!50}L & \cellcolor{grad-6!50}U \\
        \hline Amplifier NF & 5 & 7 & 7 & 5 & 6 & 8 \\
        \hline Num. of ch. & 116 & 100 & 62 & 29 & 47 & 36 \\
        \hline Num. of seg. & 15 & 6 & 4 & 2 & 3 & 2 \\
        \hline
    \end{tabular}
    \vspace{-1em}
\end{table}

To evaluate the total throughout for different optical bandwidths, we consider several scenarios where we start with a few channels in the centre of a C-band, followed by an increase in the number of channels to fully populate this band. After the C-band is populated, the L-band starts being populated with channels starting from the lowest wavelength (closer to the C-band), until the band is fully populated too. This is repeated with S-band channels from the highest wavelength, U-band from the lowest wavelength, E-band from the highest wavelength, and finally O-band from the highest wavelength. A 5~nm guard band is placed between each band. The total number of channels with fully populated bands is given in Tab.~\ref{tab:params}.

\begin{figure}[t!]
    \centering
    \clearpage{}\includegraphics{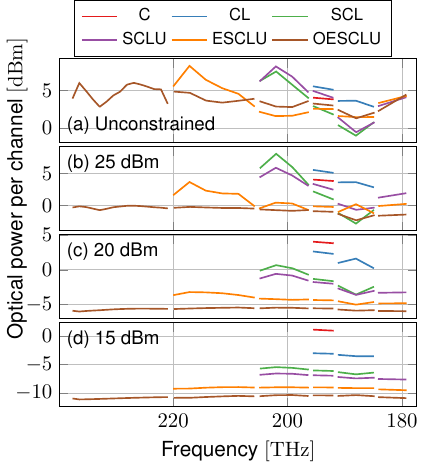}
\clearpage{}
    \vspace{-.8em}
    \caption{Optimised optical launch power for 1 span, assuming ideal transceiver, for fully populated bands. Subfigures from top to bottom show different maximum optical power constraint scenarios.}
    \label{fig:launch_power}
    \vspace{-1.8em}
\end{figure}

\begin{figure}[b!]
    \vspace{-.8em}
    \centering
    \clearpage{}\includegraphics{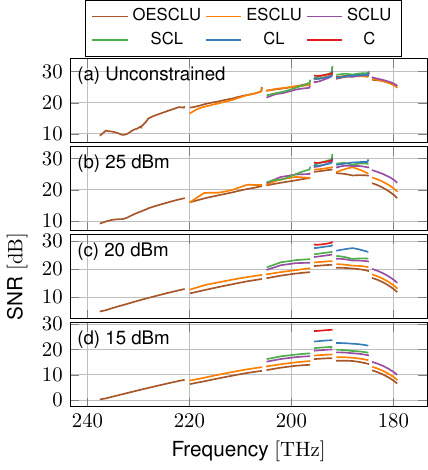}
\clearpage{}
    \caption{Signal-to-Noise Ratio for a single span, with fully populated bands, optimised assuming ideal transceiver.}
    \label{fig:snr}
\end{figure}

\section{Launch power optimisation algorithm}

For each simulated bandwidth, an optimisation of the launch power was carried out targeting maximum total system throughput for a given total power constraint $P_{_\mathrm{lim}}$. The optimisation cost function is defined in Eq.~\ref{eq:loss}.

\vspace{-1.5em}
{\small
\begin{align}
\mathcal{L} = -\sum_{i=0}^{N_\mathrm{ch}} & \log_2 \Big( 1 + 
    \frac{\tau P_i}{\eta_{\mathrm{NLI},i} (\tau P_i)^3 + P_{\mathrm{ASE},i} + P_{\mathrm{TRX}}} \Big),
\label{eq:loss}
\end{align}
}\noindent where $i$ is the channel index, $N_\mathrm{ch}$ is the number of channels, $\eta_{\mathrm{NLI},i}$ is the NLI noise coefficient, $P_{\mathrm{ASE},i}$ is the the ASE noise power, and $P_\mathrm{TRX}$ is the transceiver noise power. $P_i$ is channel launch power which is scaled by the factor $\tau$ in order to avoid exceeding the total optical power constraint $P_\mathrm{lim}$. This optimisation constraint is given by Eq.~\ref{eq:tau}.

\begin{equation}
    \tau = \min\left(1~ , ~\frac{P_\mathrm{lim}}{\sum_{i=0}^{N_\mathrm{ch}} P_i}\right) \label{eq:tau}
\end{equation}
\vspace{-1em}

\noindent The total link throughput is estimated using the Shannon capacity bound formula given by $C = -2 f_s \cdot \mathcal{L}$ with $f_s$ being the symbol rate.

The optimisation is performed employing the L-BFGS-B algorithm~\cite{2020SciPy-NMeth}. To reduce optimisation dimensionality, we divide each band signal bandwidth into segments $N_B = \mathrm{round} ( B_\mathrm{band} / B_p )$, where $B_\mathrm{band}$ is band bandwidth and $B_p$ is segment size of 750~GHz for O-Band and 1.5~THz in other bands. A maximum number of segments for fully populated bands is given in Tab.~\ref{tab:params}. The input variables for the optimiser are the segment edges, and the channel launch power is obtained using a linear interpolation between each segment edge. 

\begin{figure*}[t!]
    \centering
    \clearpage{}\includegraphics{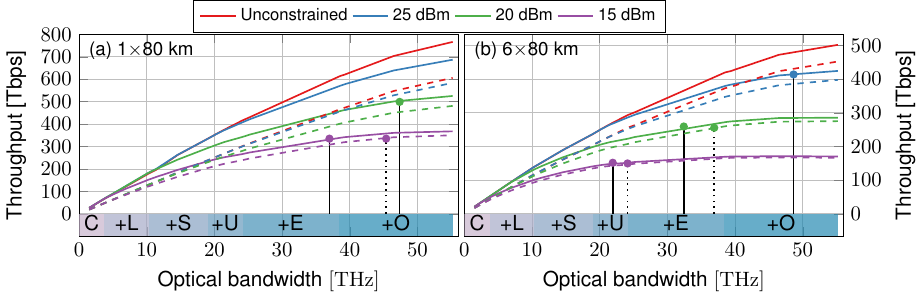}
\clearpage{}
    \vspace{-1em}
    \caption{Link throughput with an increasing number of populated channels and bands. The solid line represents simulations with ideal transceivers and dashed and dotted lines with 20~dB back-to-back SNR limited transceivers. Different colour graphs represent different total optical power constraints. (a) shows single-span simulation and (b) shows six-span simulations. Black lines show 90\% of the estimated peak throughput.}
    \label{fig:figure2}
\end{figure*}

\section{Results}

The optimal launch power allocation was obtained for all combinations of bands and is shown in Fig.~\ref{fig:launch_power} for a single-span transmission. For the unconstrained launch power case, lower wavelength bands have higher launch power due to the higher fibre attenuation within those bands combined with the ISRS effect, which transfers power from the shorter wavelength bands to the longer wavelength ones. For wavelengths within the U-band, the high-fibre attenuation also increases their optimal launch power levels. A deep spectral dip in the optimal launch power is observed in the O-band around the zero-wavelength dispersion because of the high NLI noise suffered by these channels. 

When we introduce an upper limit for the total power, the power variation in the O- and E-bands becomes less apparent. Compared to the unconstrained case, the power is reduced from the U-band when E- and O-bands are added. Similar trends can be observed for S-, C- and L- bands.

Fig.~\ref{fig:snr} shows SNR distribution over the transmission bandwidth. 
For the constrained cases, adding new bands reduces SNR in existing bands. However, for the unconstrained case, the SNR variations when adding bands are reduced, as the optimum launch power allocation can partly compensate for these variations. This latter effect shows that adding additional bands does not greatly impact already operational bands, provided that we have no constraints on the total optical power.

Fig.~\ref{fig:figure2} evaluates the maximum achievable system throughput when increasing the bandwidth for single and 6 spans and for ideal and noisy transceivers. For an ideal transceiver, the original unconstrained system with all occupied bands achieves a total throughput of 766.5~Tbps with 29.5~dBm total launch power, while imposing a 15~dBm total power limit reduces it to 367.8~Tbps. For 6-span transmission, this drop is from 502.3~Tbps to 170.6~Tbps. For a non-ideal transceiver, with 20 dB back-to-back SNR, this drop is from 606.3~Tbps to 350.3~Tbps over a single span and 452.8~Tbps to 166.7~Tbps over 6-spans. For the latter case, 90\% of the total throughput using full O-U bands was reached at 24~THz of bandwidth for 15~dBm total power limit, while with a 20~dBm power constraint throughput is saturated at a bandwidth of 36.9~THz.

\section{Conclusions}
Using a fast integral GN model, we quantified the impact of optical launch power constraints on the achievable throughput for UWB systems. Enabled by GPU parallelisation of the model, optimum launch power allocation is obtained for simulations ranging from the C-band only to the OESCLU bands.

By default, launch power optimisation does not take into account power limitations imposed due to equipment and safety regulations. Introducing the constraints in the total optical power changes the optimal power distribution, suppressing the power variations and power tilt due to SRS. 

The total achievable throughput is noticeably reduced in the presence of the total power constraint. The reduction is most marked when more transmission bands and longer transmission distances are introduced.
With a 15~dBm constraint in a single-span scenario, adding O-band to the C-to-E system results in a negligible increase in throughput, while in the 6-span scenario, adding E- and O-bands bring a negligible increase. A total power of more than 20~dBm is needed for the addition of E-band to provide an increase in throughput of more than 10\%. This figure has to be increased to 25~dBm for the further addition of O-band to offer an additional throughput increase of >10\%. 

Results show that no additional throughput is obtained by adding E- and O-bands under low total power constraints over multiple spans. 

\clearpage

\section{Acknowledgements}
This work is partly funded by the EPSRC Programme Grant TRANSNET~(EP/R035342/1) and EWOC~(EP/W015714/1). M. Jarmolovi\v{c}ius and H. Buglia are funded by the Microsoft 'Optics for the Cloud' Alliance, a UCL Faculty of Engineering Sciences Studentship, and an EPSRC studentship (EP/T517793/1).

\printbibliography
\vspace{-4mm}

\end{document}